\documentclass[reprint,groupedaddress,superscriptaddress,floatfix,prl]{revtex4-2}
\usepackage{graphicx}
\usepackage{bm}
\usepackage{dsfont}
\usepackage{amsmath, amsthm, amssymb,mathrsfs}
\usepackage{setspace}
\usepackage[utf8]{inputenc}
\usepackage[toc,page]{appendix}
\usepackage{epstopdf}
\usepackage{hyperref}
\usepackage{mdframed}
\usepackage{footnote}
\usepackage{mathtools}
\usepackage[dvipsnames]{xcolor}
\usepackage{soul}

\newcommand{\beqa}{\begin{eqnarray}}
\newcommand{\eeqa}{\end{eqnarray}}

\newcommand{\ket}[1]{\left| #1 \right\rangle}

\newcommand{\ketbra}[2]{\left|#1\right\rangle\hskip-1mm\left\langle #2\right|}

\begin{document}

\title{Exchange-Free Computation on an Unknown Qubit at a Distance}
\author{Hatim Salih}
\email{salih.hatim@gmail.com}
\affiliation{Quantum Engineering Technology Laboratory, Department of Electrical and Electronic Engineering, University of Bristol, Woodland Road, Bristol, BS8 1UB, UK}
\affiliation{Quantum Technology Enterprise Centre, HH Wills Physics Laboratory, University of Bristol, Tyndall Avenue, Bristol, BS8 1TL, UK}
\author{Jonte R. Hance}
\affiliation{Quantum Engineering Technology Laboratory, Department of Electrical and Electronic Engineering, University of Bristol, Woodland Road, Bristol, BS8 1UB, UK}
\author{Will McCutcheon}
\affiliation{Quantum Engineering Technology Laboratory, Department of Electrical and Electronic Engineering, University of Bristol, Woodland Road, Bristol, BS8 1UB, UK}
\affiliation{Institute of Photonics and Quantum Science, School of Engineering and Physical Sciences, Heriot-Watt University, Edinburgh, EH14 4AS, UK}
\author{Terry Rudolph}
\affiliation{Department of Physics, Imperial College London, Prince Consort Road, London SW7 2AZ, United Kingdom}
\author{John Rarity}
\affiliation{Quantum Engineering Technology Laboratory, Department of Electrical and Electronic Engineering, University of Bristol, Woodland Road, Bristol, BS8 1UB, UK}
\date{\today}

\begin{abstract}
We present a way of directly manipulating an arbitrary qubit, without the exchange of any particles. This includes as an application the exchange-free preparation of an arbitrary quantum state at Alice by a remote classical Bob. As a result, we are able to propose a protocol that allows one party to directly enact, by means of a suitable program, any computation exchange-free on a remote second party's unknown qubit. Further, we show how to use this for the exchange-free control of a universal two-qubit gate, thus opening the possibility of directly enacting any desired algorithm remotely on a programmable quantum circuit.

\end{abstract}

\maketitle


Quantum physics opens up the surprising possibility of obtaining knowledge from, or through, places where no information-carrying particles have been. This was first proposed and subsequently demonstrated experimentally in the context of computing \cite{Mitchison2001CFComputation, Hosten2006CounterComp}, where the result of a computation is learnt based on the phenomena of interaction-free measurement and the Zeno effect \cite{Elitzur1993Bomb, Kwiat1995IFM, kwiat1999high, Misra1977Zeno, Rudolph2000Zeno}. More specifically, without any photons entering or leaving an optical circuit, the result of a computation is obtained without the computer ever `running'.

Just as intriguing was the proposal and subsequent experimental demonstration of a simple quantum scheme for allowing two remote parties to share a cryptographic random bit-string, without exchanging any information-carrying particles \cite{Noh2009CounterfactualCrypto,Liu2012NohDemons}. The fact that the protocol had limited maximum-efficiency was not a serious a drawback for its purpose since the shared information was random, meaning failed attempts could simply be discarded in the end. This, however, begged the question whether efficient, deterministic communication was possible exchange-free, that is without particles crossing the communication channel. 

In 2013, building on the ideas above, Salih et al devised a scheme allowing two remote parties to efficiently and deterministically share a message exchange-free, in the limit of a large number of protocol cycles and ideal practical implementation \cite{Salih2013Protocol}. The protocol was recently demonstrated experimentally by Pan and colleagues \cite{Pan2017Experiment}. Importantly, the previously-heated debate over whether the laws of physics even allow such communication (for both bit values) seems to be settling; Nature does allow exchange-free communication (and therefore computation) \cite{Vaidman2014SalihCommProtocol,Salih2014ReplyVaidmanComment,Griffith2016Path,Salih2018CommentPath,Griffiths2018Reply,Aharonov2019Modification, Salih2018Laws}.

We present in what follows a protocol allowing a remote Bob to prepare any qubit he wishes at Alice without any particles passing between them, thus exchange-free. This is different from counterfactually sending a quantum state from Bob to Alice by means of counterportation \cite{Salih2016Qubit,*Salih2014Qubit,Salih2018Paradox}, in that Bob does not need to prepare a quantum object at his end (a quantum superposition of blocking and not blocking the optical communication channel) thus making the scheme much easier to implement. More generally, Bob can directly apply any arbitrary Bloch-sphere rotation to an unknown qubit at Alice---in other words, any single-qubit quantum computation. Note that we use ``exchange-free" and ``counterfactual" interchangeably. While we describe an optical realisation using photon polarisation, the scheme is in principle applicable to other physical implementations---and helps advance quantum information science.

\begin{figure*}
    \centering
    \includegraphics[width=0.6\linewidth]{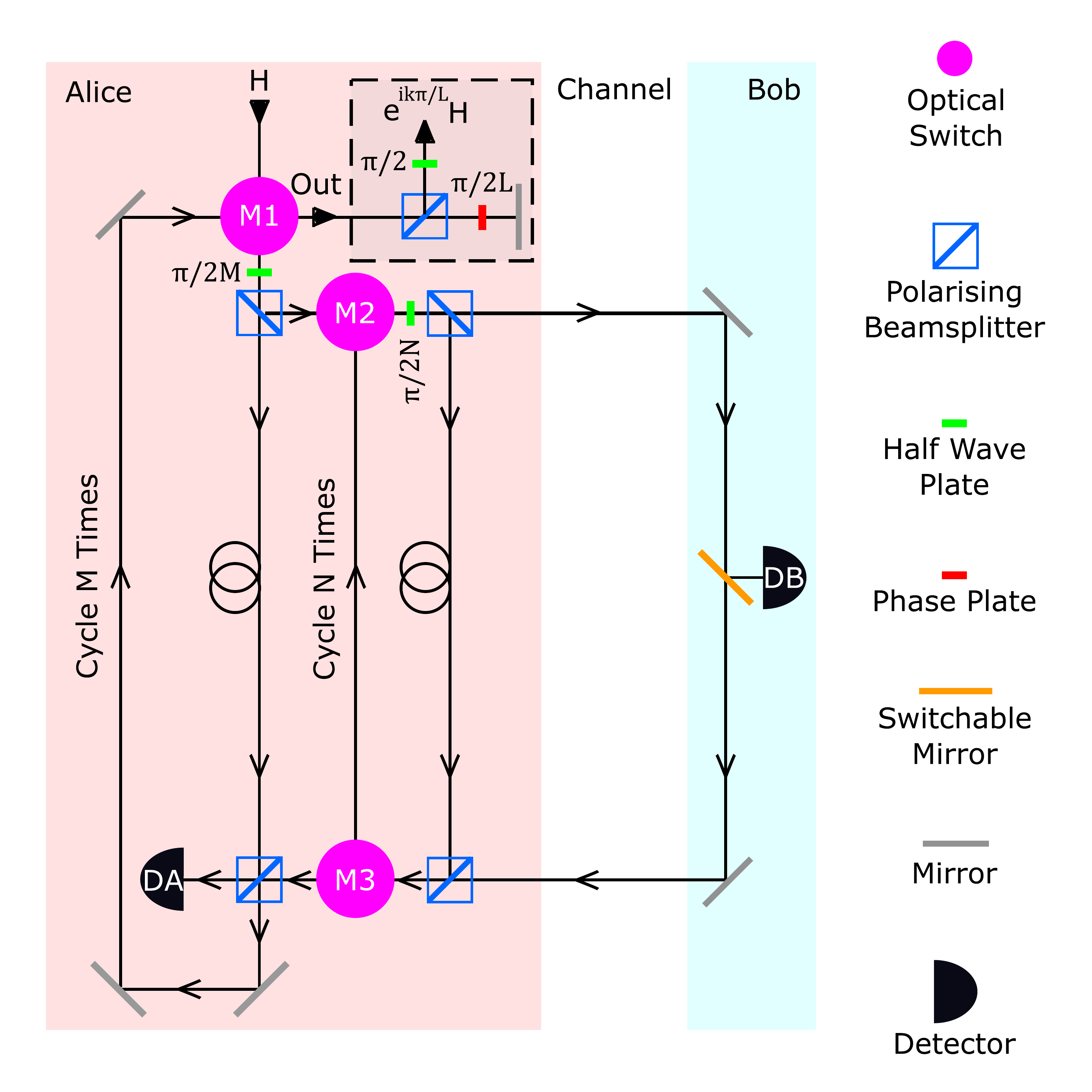}
    \caption{Our exchange-free Phase Unit, which applies a phase determined by Bob to Alice's $H$-polarised input photon. The Phase Unit comprises an equivalent setup to that of Salih et al's 2013 protocol \cite{Salih2013Protocol}, but with an added phase-module in the dashed box. The optical switches each alter the paths at different times in the protocol to allow the photon to do the correct number of cycles. Optical switch $M1$ inserts the photon into the device, and keeps it in for $M$ outer cycles; optical switches $M2$ and $M3$ cycle the photon around for $N$ inner cycles per outer cycle. The Polarising Beamsplitters transmit $H$-polarised light, and reflect $V$-polarised light. The half wave plates are tuned to implement $\hat{\textbf{R}}_y(\theta)$ rotations on polarisation with $\theta$ of $\pi$, $\pi/M$ and $\pi/N$, as shown in the figure. As explained in the text, detectors $D_A$ and $D_B$ not clicking ensure that the photon has not been to Bob. After $M$ outer cycles, the photon is sent by $M1$ to the right. The photon only exits the Phase Unit if its polarisation had been flipped to $V$ as a result of Bob blocking the channel (which he does by switching his Switchable Mirror on) because of the action of the Polarising Beamsplitter in the dashed box. The phase plate (tuned to enact a $\hat{\textbf{R}}_z$, or phase, rotation, and realisable using a tiltable glass plate) adds a phase of $\pi/2L$ to the photon every time it passes through it, summing to $\pi/L$ every time it is sent $H$-polarised to the right by M1. Bob doesn't block for $k$ runs (out of a maximum $L$), then blocks, allowing him to set the final phase of the photon, $k\pi/L$, anywhere from $0$ to $\pi$, in increments of $\pi/L$. An initially $V$-polarised photon can be put through an altered version of this device to add a phase to it (identical, except for the $\pi/2$ half-wave plate being moved to above $M1$). The unit rotates Alice's qubit by $\hat{\textbf{R}}_z(k\pi/L)$.}
    \label{fig:phaseunit}
\end{figure*}
\begin{figure}
    \centering
    \includegraphics[width=\linewidth]{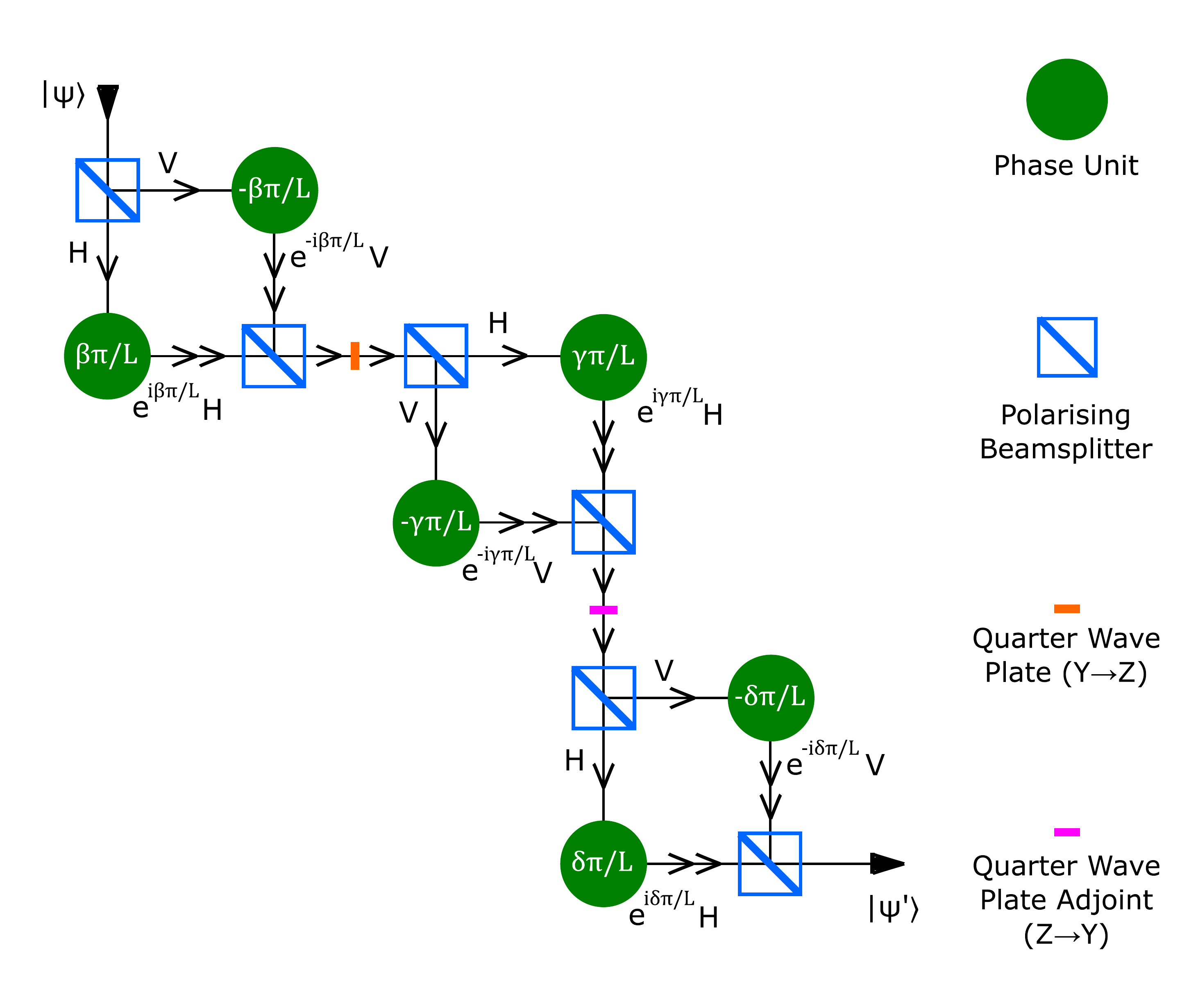}
    \caption{The overall protocol, incorporating multiple Phase Units from Fig.\ref{fig:phaseunit}, as well as polarising beamsplitters (which transmit horizontally-polarised, and reflect vertically-polarised, light), as well as a quarter wave plate and its adjoint (conjugate-transpose). The setup allows Bob to implement any arbitrary unitary on any initial pure state $\ket{\psi}$ Alice inserts, entirely exchange-free.}
    \label{fig:OverallProt}
\end{figure}

Our protocol consists of a number of nested outer interferometers, each containing a number of inner interferometers, as in Salih et al's 2013 protocol \cite{Salih2013Protocol}. We combine these interferometers into a device that we call a Phase Unit, allowing Bob to apply a relative phase to Alice's photonic qubit (Fig.\ref{fig:phaseunit}). We pair two Phase Units such that one applies some phase to Alice's $H$-polarised component, while the other applies an equal but opposite-sign phase to her $V$-polarised component, resulting in a $\hat{\textbf{R}}_z(\theta)$ rotator. By chaining three such $\hat{\textbf{R}}_z(\theta)$ rotators, interspersed with appropriate wave-plates, Bob can apply any arbitrary unitary to Alice's qubit, exchange-free (Fig.\ref{fig:OverallProt}).

Note, we define the Bloch sphere for polarisation such that the poles are $\ket{H}$ and $\ket{V}$, and the rotations are
\begin{equation}
        \hat{\textbf{R}}_x(\theta) =
        \begin{pmatrix}
        \cos{\big(\frac{\theta}{2}\big)}  &  -i\sin{\big(\frac{\theta}{2}\big)}\\
        i\sin{\big(\frac{\theta}{2}\big)}  &  \cos{\big(\frac{\theta}{2}\big)}
        \end{pmatrix}
        = e^{-i\theta\hat{\sigma}_x/2}
        \end{equation}
        \begin{equation}
        \hat{\textbf{R}}_y(\theta) =
        \begin{pmatrix}
        \cos{\big(\frac{\theta}{2}\big)}  &  -\sin{\big(\frac{\theta}{2}\big)}\\
        \sin{\big(\frac{\theta}{2}\big)}  &  \cos{\big(\frac{\theta}{2}\big)}
        \end{pmatrix}
        = e^{-i\theta\hat{\sigma}_y/2}
        \end{equation}
        \begin{equation}
        \hat{\textbf{R}}_z(\theta) =
        \begin{pmatrix}
    e^{-i\theta/2}  &  0\\
    0  &  e^{i\theta/2}
    \end{pmatrix}
     = e^{-i\theta\hat{\sigma}_z/2}
\end{equation}
for dummy variable $\theta$, and Pauli matrices $\hat{\sigma}_{x,y,z}$.

We first go through Salih et al's 2013 protocol. However, we describe the protocol, following \cite{Salih2018Paradox}, without any reference to either interaction-free measurement or the Zeno effect of \cite{Misra1977Zeno,Elitzur1993Bomb}. In order to do this, we think of our detectors as being placed far enough, such that they perform no measurement before the photon had had time to exit the protocol. Any photonic component travelling towards either detector can thus be thought of as entering a loss mode, meaning that if the photon exits the protocol successfully then it cannot have taken the path towards that detector, and the detector will subsequently not register a click.

To start with, a photon of state $a\ket{H}+b\ket{V}$ enters the outer interferometer through a half wave plate (HWP) tuned to apply a $\hat{\textbf{R}}_y(\pi/M)$ rotation. The photon then enters a polarising beam splitter (PBS), which transmits horizontal polarisation, but reflects vertical polarisation.

The $V$-polarised component circles through a series of $N$ inner interferometers, where, in each, it goes through a HWP tuned to apply a $\hat{\textbf{R}}_y(\pi/N)$ rotation, then through another PBS. The $H$-polarised component from this PBS passes across the channel, from Alice to Bob, who can choose to block or not block, by switching on or off his switchable mirror. If he blocks, this $H$-polarised component goes into a loss mode towards detector $D_B$; if not, it returns to Alice's side, recombines at another PBS with the $V$-polarised component, then enters the next inner interferometer. After the chain of $N$ inner interferometers, the resulting components are then passed through one final PBS, sending any $H$-polarised component that has been to Bob into a loss mode towards detector $D_A$, before being recombined at another PBS with the $H$-polarised component from the arm of the outer interferometer. Importantly, neither detector clicking, ensures that the photon has not been to Bob. 

As each inner interferometer applies $\hat{\textbf{R}}_y(\pi/N)$, if Bob doesn't block, the rotations sum to
\begin{equation}
    \begin{split}
       \hat{\textbf{U}}_{NB}^N=(e^{-i\pi\hat{\sigma}_y/2N})^N
    =e^{-i\pi\hat{\sigma}_y/2} = \hat{\textbf{R}}_y(\pi)
    \end{split}
\end{equation}

Therefore, the state after the inner interferometer chain is
\begin{equation}
    \begin{split}
       \ket{V}_I\rightarrow\hat{\textbf{U}}_{NB}^N\ket{V}_I=\ket{H}_I\rightarrow Loss
    \end{split}
\end{equation}

This means the $V$-polarised component becomes $H$-polarised, entering the loss mode towards detector $D_A$ after the final PBS, meaning the only component of the wavefunction exiting the outer interferometer is the $H$-polarised one that went via the outer arm.

Similarly, if Bob blocks for all inner interferometers,
\begin{equation}
    \begin{split}
       \hat{\textbf{A}}_{B}^N=&\Bigg[
       e^{-i\pi\hat{\sigma}_y/2N}
        \begin{pmatrix}
        1  &  0\\
        0  &  0
    \end{pmatrix}\Bigg]^N\\
        =&
         \begin{pmatrix}
       \cos{(\frac{\pi}{2N})}^N  &   0\\
       \cos{(\frac{\pi}{2N})}^{N-1}\sin{(\frac{\pi}{2N})}  &   0
        \end{pmatrix}
    \end{split}
\end{equation}

Therefore, the state after an outer interferometer is
\begin{equation}
    \begin{split}
       \ket{V}_I\rightarrow
       &\hat{\textbf{A}}_{B}^N\ket{V}_I\\
       =&\cos{\big(\frac{\pi}{2N}\big)}^N\ket{V}_I+\cos{\big(\frac{\pi}{2N}\big)}^{N-1}\sin{\big(\frac{\pi}{2N}\big)}\ket{H}_I\\
       \rightarrow& \cos{\big(\frac{\pi}{2N}\big)}^N\ket{V}+Loss
    \end{split}
\end{equation}
meaning some $V$-polarised component exits the outer interferometer. 

If Bob, doesn't block, the outer cycle applies
\begin{equation}
    \begin{split}
    \begin{pmatrix}
     1  &  0\\
     0  &  0
\end{pmatrix}
e^{-i\pi\hat{\sigma}_y/2M}
    \end{split}
\end{equation}

If he does block, the outer cycle applies
\begin{equation}
\begin{split}
\begin{pmatrix}
     1  &  0\\
     0  &  \cos{\big(\frac{\pi}{2N}\big)}^N
\end{pmatrix}
e^{-i\pi\hat{\sigma}_y/2M}
\end{split}
\end{equation}

We repeat this $M$ times, starting with a $H$-polarised photon, and using a final PBS to split it into $H$- and $V$-polarised components.

As Alice applies a $\hat{\textbf{R}}_y(\pi/M)$ rotation at the start of each outer interferometer, if Bob doesn't block, the state of the photon after $M$ outer cycles is
\begin{equation}
    \begin{split}
    \cos{\big(\frac{\pi}{2M}\big)}^M\ket{H}
    \end{split}
\end{equation}

Therefore, if the photon isn't lost, it remains $H$-polarised. However, if Bob blocks, the photon after $M$ outer cycles (as $N\rightarrow\infty$) becomes $V$-polarised.

To prepare any qubit at Alice, Bob needs to apply a relative phase between Alice's two component, which can be represented as a $\hat{\textbf{R}}_z(\theta)$ rotation. Bob can implement this exchange-free using the device in Fig.\ref{fig:phaseunit}, for an $H$-polarised component, relative to some other $V$-polarised component (e.g. one separated beforehand using a polarising beamsplitter).

We put this $H$-polarised component through one run of Salih et al's 2013 protocol, with Bob either always blocking or not blocking his channel. If he blocks, and the component exits $V$-polarised, the PBS sends it through a half wave plate that flips it to $H$-polarised, and it is kicked out of the device; however, if it is $H$-polarised, it goes through a phase plate (gaining a phase increase of $\pi/2L$), hits a mirror, goes back through the phase plate (gaining another phase increase of $\pi/2L$, for a total increase of $\pi/L$), and re-enters the device for another run.

This is repeated $L$ times, with Bob blocking or not blocking for all outer cycles in a given run. After each run, the component goes into a PBS: if it is $H$-polarised, it gains a phase of $\pi/L$; if $V$-polarised, it is flipped to $H$-polarised and sent out from the unit. Bob first doesn't block for $k$ runs, applying a phase of $k\pi/L$, then blocks, applying the transformation
\begin{equation}
    \ket{H} \rightarrow e^{ik\pi/L}\ket{H}
\end{equation}

When $N$ is finite, the rotations applied by each outer cycle when Bob blocks are not complete, meaning one run ($M$ outer cycles) doesn't fully rotate the state from $H$ to $V$. However, given Bob only blocks after the component has had a phase applied to it, to kick the component out of the device, any erroneous $H$-polarised component can be kept in the device by Bob not blocking for the remaining $L-k$ full runs afterwards, letting us treat the erroneous $H$-component as loss.

While coarse-grained for finite $L$, as $L$ goes to infinity (with $0\leq k/L\leq1$), Bob can generate any relative phase for Alice's qubit, from $0$ to $\pi$. Further, by moving the $\pi/2$ half-wave plate from its location in Fig.\ref{fig:phaseunit} to the input, a similar phase can be added to a $V$-polarised component, relative to a $H$-polarised component.

Moreover, the Phase Unit can be constructed to include Aharonov and Vaidman's clever modification of Salih et al's 2013 protocol \cite{Aharonov2019Modification}, satisfying their weak-measurement criterion for exchange-free communication. We do this by running the inner cycles for $2N$ cycles rather than $N$, except that for the case of Bob not blocking, he instead blocks for one of the $2N$ inner cycles, namely the $N$th inner cycle. This has the effect of helping to remove any lingering $V$ component exiting the inner interferometer of Fig.\ref{fig:phaseunit} due to imperfections in practical implementation.       

We now use our Phase Unit as the building block for a protocol where Bob can implement any arbitrary unitary onto Alice's qubit, exchange-free.

Any arbitrary $2\times2$ unitary matrix can be written as
\begin{equation}
\begin{split}
    \hat{\textbf{U}} 
     &=e^{i(2\alpha'-\beta'\hat{\sigma}_z-\gamma'\hat{\sigma}_y-\delta'\hat{\sigma}_z)/2}\\
     &=e^{i\alpha'} \hat{\textbf{R}}_z(\beta') \hat{\textbf{R}}_y(\gamma') \hat{\textbf{R}}_z(\delta')
\end{split}
\end{equation}

Note, the factor of $e^{i\alpha'}$ can be ignored, as it provides global rather than relative phase, which is unphysical for a quantum state \cite{nielsenchuang2002}.

We can apply the $\hat{\textbf{R}}_z(\theta)$ rotations using the Phase Unit, and make a $\hat{\textbf{R}}_y(\theta)$ rotation by sandwiching a $\hat{\textbf{R}}_z(\theta)$ rotation between a $-\pi/4$-aligned Quarter Wave Plate, $\hat{\textbf{U}}_{QWP}$, and its adjoint, $\hat{\textbf{U}}^{\dagger}_{QWP}$, where
\begin{equation}
    \begin{split}
        \hat{\textbf{U}}_{QWP} =& 
   \hat{\textbf{R}}_x(-\pi/2)
     = e^{i\pi\hat{\sigma}_x/4}\\
     \hat{\textbf{U}}^{\dagger}_{QWP} =&
   \hat{\textbf{R}}_x(\pi/2)
    = e^{-i\pi\hat{\sigma}_x/4}
    \end{split}
\end{equation}

We set 
\begin{equation}
    \begin{split}
        \beta' = 2\pi\beta/L,\;
        \gamma' = 2\pi\gamma/L,\;
        \delta' = 2\pi\delta/L
    \end{split}
\end{equation}
where, for the three Phase Unit runs, $k$ is $\beta,\;\gamma$ and $\delta$.

The Phase Units form components of the overall protocol, as shown in Fig.\ref{fig:OverallProt}. Here, Alice first splits her input state $\ket{\psi}$ into $H$- and $V$-polarised components with a polarising beamsplitter (PBS), before putting each component through a Phase Unit, to generate equal and opposite phases on each. She recombines these at another PBS. Afterwards, she puts the components through a quarter wave plate, then through another run of PBS, Phase Unit, and PBS, then through the conjugate-transpose of the quarter-wave plate, tuned to convert the partial $\hat{\textbf{R}}_z$ rotation (phase rotation) into a partial $\hat{\textbf{R}}_y$ rotation. Finally, she applies another run of PBS, Phase Unit, and PBS to implement a second $\hat{\textbf{R}}_z$ rotation.

Using these $\hat{\textbf{R}}_z$ and $\hat{\textbf{R}}_y$ rotations, Bob can implement any arbitrary rotation on the surface of the Bloch sphere on Alice's state. This can be used either to allow Bob to prepare an arbitrary pure state at Alice (if she inserts a known state, such as $\ket{H}$), or to perform any arbitrary unitary transformation on Alice's qubit, without Bob necessarily knowing that input state.

Because the Phase Units output their respective photon components after Bob blocks for a run, the timing of which depends on the phase Bob wants to apply, there is a time-binning (a grouping of exit times into discrete bins) of the components from each Phase Unit correlated with the phase Bob applies in that unit. Bob can, on his side, compensate for the time-binning (given he knows the phase he is applying). Further, in order to locate the photon in time, Alice can detect the time of exit using a non-demolition single photon detector.

Alternatively, we could add a final pair of Phase Units with the value of $k$ set to $3L'-\beta-\gamma-\delta$ (where $L'$ is the value of $L$ for each of the first three Phase Unit pairs, and $\beta$, $\gamma$ and $\delta$ are their respective $k$-values), but without phase plates (see Fig.\ref{fig:phaseunit}). This means that while no phase is applied, a time delay is still added to the components, meaning the photon always exits the overall device at a time proportional to $3L$, rather than $\beta$, $\gamma$ and/or $\delta$ as before. This makes the time of exit uncorrelated to Bob's unitary, which means Alice can know in advance the expected exit time of her photon from the protocol (without needing to perform a non-demolition measurement to find it).

\begin{figure}
    \centering
    \includegraphics[width=\linewidth]{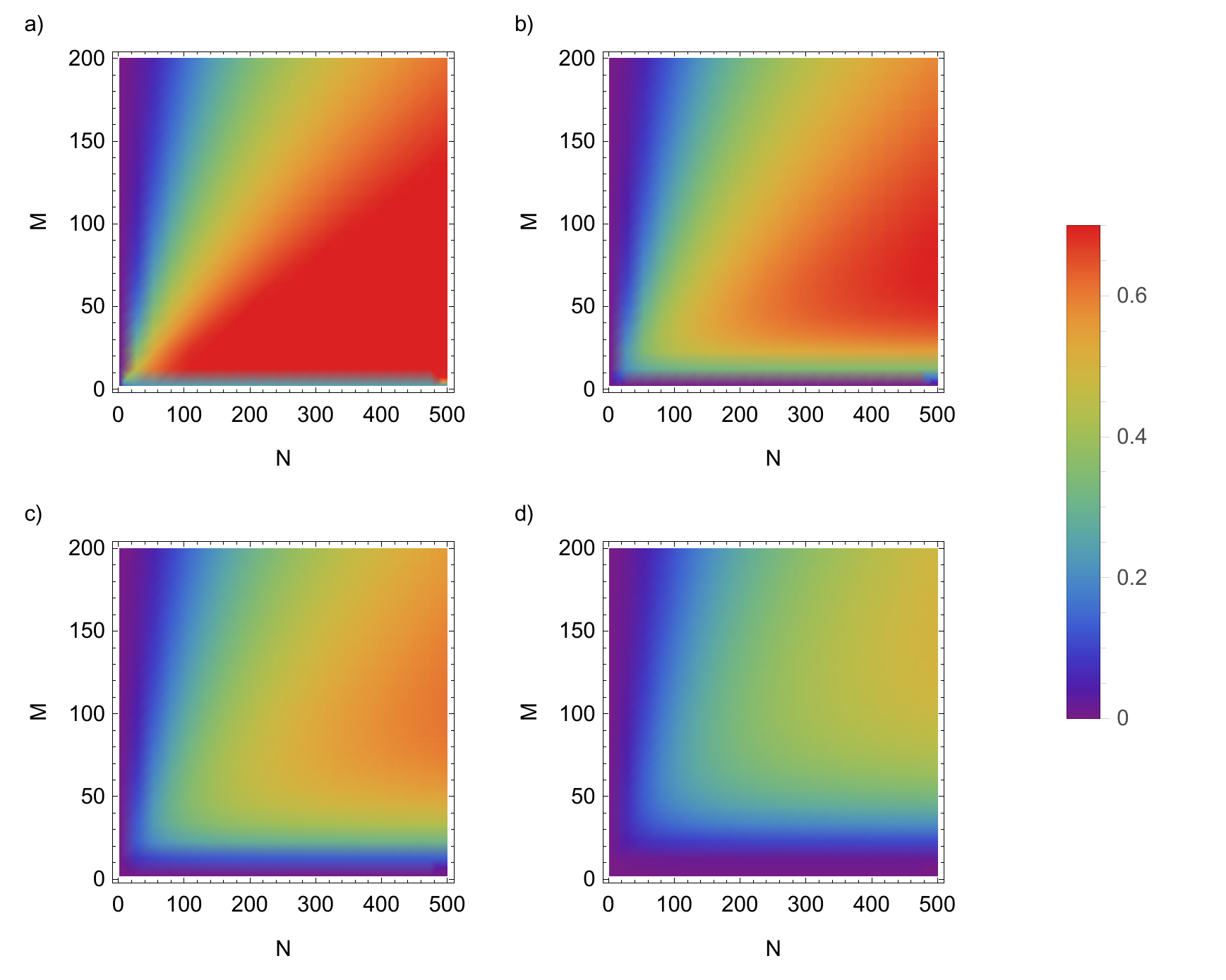}
    \caption{The survival probability of a photon going through a Phase Unit (Fig.\ref{fig:phaseunit}) of given M (number of outer cycles) and N (number of inner cycles). This is shown for the unit imparting phase $i k \pi/L$, where $k$, the number of runs of the protocol before the photon is emitted from the unit, is 1 for (a), 5 for (b), 10 for (c) and 20 for (d). Note there is no dependence on $L$, the maximum number of runs.}
    \label{fig:PSurvPhaseUnit}
\end{figure}

When considering a finite number of outer and inner cycles, there is a nonzero probability of the photon not returning to Alice, which reduces the protocol's efficiency. The survival probability of a photon going through a Phase Unit is plotted in Fig.\ref{fig:PSurvPhaseUnit}. The survival probability for the overall protocol is the product of the survival probability for the three Phase Units:
\begin{equation}
    P(Tot)_{Sv} = P(\beta)_{Sv}\cdot P(\gamma)_{Sv}\cdot P(\delta)_{Sv}
\end{equation}

As expected, as $\{M,N\}\rightarrow\infty$, the survival probability goes to one.


Regardless, postselection renormalises Alice's output state such that if Alice receives an output photon, it will be in a pure state. Thus, for our set-up, given ideal optical components, the rotation enacted on Alice's qubit is always the rotation Bob has applied, not just for any $L$, but also for any $N$, $M$, and $k$.

Interestingly, a Phase Unit, which outputs a photon into one of $L$ different time bins depending on the number of runs Bob blocks, could be adapted to sending, exchange-free, a classical logical state of dimension $d$ greater than two---a ``dit", rather than a bit. We do this by removing the phase plate in the Phase Unit (see Fig.\ref{fig:phaseunit}). Bob first doesn't block for $k$ runs, then blocks for the remaining $L-k$, meaning the photon's output occurs in the $k^{th}$ time-bin of $L$ possible time-bins. This encodes a dit of dimension $L$ into the photon, which Alice can read via non-demolition single photon detector (NDSPD).

\begin{figure}
    \centering
    \includegraphics[width=\linewidth]{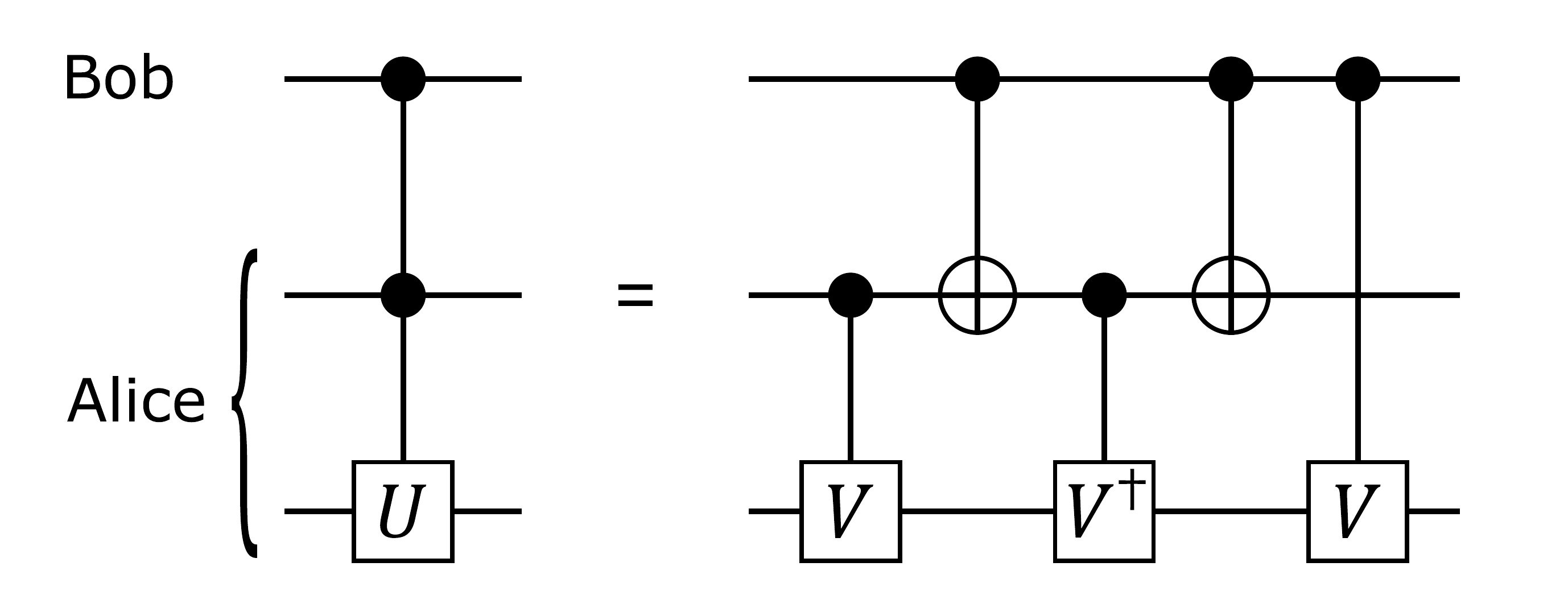}
    \caption{Quantum circuit diagram, showing how a 3-qubit gate applying a controlled-controlled unitary $U$ can be constructed from two-qubit gates along with single-qubit gates, where $U$ is some unitary transformation, and $V^2=U$ \cite{Barenco1995Gates}. Using our exchange-free single-qubit gate, a classical Bob can directly simulate the control action on Alice's photonic qubits. Since any quantum circuit can be constructed using 2-qubit gates along with single-qubit ones, our exchange-free single-qubit gate allows Bob in principle to directly program any quantum algorithm at Alice, without exchanging any photons.}
    \label{fig:QCircuit}
\end{figure}

We now show how our exchange-free protocol enabling arbitrary single-qubit operations, can in principle allow a classical Bob to directly enact any quantum algorithm he wishes on Alice's qubits, without exchanging any particles with her. This is based on the fact any quantum algorithm can be efficiently constructed from 2-qubit operations (such as CNOT) and single-qubit ones. Our protocol already enables exchange-free single-qubit operations, i.e. gates. Thus, if Bob can directly activate or not, a 2-qubit gate at Alice, exchange-free, then directly programming an entire quantum algorithm at Alice using these two building blocks becomes possible. The quantum network of figure Fig.\ref{fig:QCircuit} shows how a 3-qubit controlled-controlled gate, applying some unitary $U$ to the target qubit at Alice, can be constructed from 2-qubit controlled gates \cite{Barenco1995Gates}. (Some controlled-controlled gates can have more implementable circuits than the general one given here \cite{nielsenchuang2002}.) A classical Bob, at the top end of Fig.\ref{fig:QCircuit}, uses our exchange-free single-qubit gate to simulate the control action on Alice's photonic qubits. For simulating the CNOT gate, he can choose to either apply the identity transformation, representing control-bit $\ket{0}$, or apply an $X$ transformation, representing control-bit $\ket{1}$. For the controlled-$V$ gate, he can choose to either apply the identity transformation, again representing control-bit $\ket{0}$, or apply a $V$ transformation, representing control-bit $\ket{1}$. In this scenario, we envisage an optical programmable circuit, with exchange-free single-qubit gates acting on Alice's qubits that Bob can directly program, and 2-qubit gates acting on Alice's qubits that he can directly choose to activate, exchange-free.

In summary, we have presented a protocol allowing Bob to directly perform any computation on a remote Alice's qubit, without exchanging any photons between them. We use this to show how, in principle, Bob can directly enact any quantum algorithm at Alice, exchange-free.

\textit{Acknowledgements---} We thank Claudio Marinelli for helping to bring this collaboration together. This work was supported by the Engineering and Physical Sciences Research Council (Grants EP/P510269/1, EP/T001011/1, EP/R513386/1, EP/M013472/1 and EP/L024020/1).

\bibliography{ref.bib}

\appendix

\section{A Simpler Protocol for applying an \texorpdfstring{$\hat{\textbf{R}}_y$}{Ry} Rotation}

While considering how Bob could prepare an arbitrary qubit at Alice exchange-free, we've stumbled upon a much simpler protocol for Bob to prepare exchange-free a qubit with real, positive probability amplitudes.

Consider Fig.\ref{fig:phaseunit}, without the phase module. Starting with Alice's $H$-polarised photon, instead of Bob blocking or not blocking every cycle, he instead doesn't block for the first $M-k$ outer cycles, then blocks for the rest. In order to eliminate the error resulting from a finite number of blocked inner cycles, Alice introduces loss, attenuating the outer arm of the interferometer on her side by a factor of $\cos{(\pi/2N)}^N$ for each outer cycle. This means, before the final PBS, the state is
\begin{equation}
   \begin{split}
     \ket{\Psi}=&\cos{\big(\frac{\pi}{2N}\big)}^{MN}\cos{\big(\frac{\pi}{2M}\big)}^{M-k}\\
     &\Big(\cos{\big(\frac{k\pi}{2M}\big)}\ket{H}+\sin{\big(\frac{k\pi}{2M}\big)}\ket{V}\Big)
  \end{split}
\end{equation}

By postselecting on Alice's photon successfully exiting the protocol, she receives the state
\begin{equation}
  \begin{split}
\ket{\Psi}_{PS}=\cos{\big(\frac{k\pi}{2M}\big)}\ket{H}+\sin{\big(\frac{k\pi}{2M}\big)}\ket{V}
\end{split}
\end{equation}

By choosing $k$, Bob directly applies a $\hat{\textbf{R}}_y$ rotation to Alice's $\ket{H}$ input state. Now, in order to allow Bob to apply such a rotation to an arbitrary input polarisation state, Alice's photon is initially split into $H$ and $V$-components using a PBS. The desired rotation is applied separately. In the case of the $V$-component, its polarisation is first flipped to $H$ before the rotation is applied, followed by a phase flip and a polarisation flip upon exit. The separate components can then be combined using a 50:50 beamsplitter, with the correct state obtained 50\% of the time. The advantage, however, is that, assuming perfect optical components and a large number of cycles, only two runs of the protocol are needed on average.

\section{Kraus Operator Notation}

Viewing exchange-free communication more abstractly, we consider the communication channel in Kraus operator notation.

Here, we associate a channel $\mathcal{X}$ to the set of Kraus operators $\{X_i\}_i$ which describe its action on a given density operator such that
\begin{equation}
\begin{split}
\mathcal{X} \sim \{ X_i\}_i\\
\rho \rightarrow \mathcal{X}(\rho)= \sum_i X_i \rho X_i^\dagger\\
\sum_i X^\dagger_i X_i = \mathds{1}
\end{split}
\end{equation}
In general, for channels $\mathcal{X}$ and $\mathcal{Y}$, their composition can be written,
\begin{equation}
\begin{split}
\mathcal{X}\circ \mathcal{Y} (\rho) := \mathcal{X}(\mathcal{Y}(\rho)) &= \sum_i X_i \bigl(\sum_j Y_j\rho Y_j^\dagger\bigr) X_i^\dagger\\
\end{split}
\end{equation}
and we denote the $N$-fold composition of a channel $\mathcal{X}^N(\rho):= \mathcal{X}\circ\mathcal{X}\circ\dots\circ\mathcal{X}(\rho)$.

In this manner we can define three channels in this protocol: first that constituting Bob's action on the channel, $b$, that goes via him, when he blocks/doesn't block
\begin{equation}
\begin{split}
\mathcal{B}^{B}\sim\{\ketbra{0_b}{1_b},\ketbra{0_b}{0_b}\},\\
\mathcal{B}^{NB}\sim\{\ketbra{1_b}{1_b},\ketbra{0_b}{0_b}\};
\end{split}
\end{equation}
Each cycle of Alice's inner-interferometer is given by $\mathcal{B}^{NB}\circ \hat{\textbf{R}}_{y,(a1,b)}^{(\pi/N)}$, and imposing that the initial state in Bob's mode is vacuum, and omitting it from the output state by tracing it out, we have that over the $N$ inner cycles one finds channels on Alice's mode $a1$ given by,
\begin{equation}
\begin{split}
\mathcal{A}_1^{B}(\rho)&:= \text{tr}_b \bigl[ (\mathcal{B}^{B}\circ \hat{\textbf{R}}_{y,(a1,b)}^{(\pi/N)})^N(\rho \otimes \ketbra{0_b}{0_b})\bigr],\\
\mathcal{A}_1^{NB}(\rho)&:= \text{tr}_b \bigl[ (\mathcal{B}^{NB}\circ \hat{\textbf{R}}_{y,(a1,b)}^{(\pi/N)})^N(\rho \otimes \ketbra{0_b}{0_b})\bigr],
\end{split}
\end{equation}
where any channel acting on a larger Hilbert space than that on which it's defined acts as identity channel, i.e. $\mathcal{B}^{NB}\sim \mathcal{B}^{NB}\otimes\mathds{1}$. We find when Bob blocks/doesn't block:
\begin{equation}
\begin{split}
\mathcal{A}_1^{B}\sim&\{
\cos{(\frac{\pi}{2N})}^N\ketbra{1_{a1}}{1_{a1}},
\ketbra{0_{a1}}{0_{a1}},\\
&\cos{(\frac{\pi}{2N})}^{N-1}\sin{(\frac{\pi}{2N})}\ketbra{0_{a1}}{1_{a1}}\},\\
\mathcal{A}_1^{NB}\sim&\{
\ketbra{0_{a1}}{1_{a1}},
\ketbra{0_{a1}}{0_{a1}}\};
\end{split}
\end{equation}
then finally the effect this has overall as the channel created by a chain of $M$ outer interferometers on Alice's inner and outer interferometer ($V$ and $H$) modes, when Bob blocks/doesn't block:
\begin{equation}
\begin{split}
\mathcal{A}_{12}^{B}(\rho)&:= (\mathcal{A}_1^{B}\circ \hat{\textbf{R}}_{y,(a2,a1)}^{(\pi/M)})^M(\rho),\\
\mathcal{A}_{12}^{NB}(\rho)&:= (\mathcal{A}_1^{NB}\circ \hat{\textbf{R}}_{y,(a2,a1)}^{(\pi/M)})^M(\rho)
\end{split}
\end{equation}

Therefore, we find
\begin{equation}
\begin{split}
\mathcal{A}_{12}^{B}\sim \Bigg\{
c_1&\ketbra{1_{a2}0_{a1}}{1_{a2}0_{a1}},\\
c_2\ketbra{0_{a2}1_{a1}}{0_{a2}1_{a1}},
c_3&\ketbra{0_{a2} 1_{a1}}{1_{a2} 0_{a1}},\\
c_4\ketbra{1_{a2} 0_{a1}}{0_{a2} 1_{a1}},
&\ketbra{0_{a2}0_{a1}}{0_{a2}0_{a1}}\\
\sqrt{(1-c_1^2-c_3^2)}&\ketbra{0_{a2}0_{a1}}{1_{a2}0_{a1}},\\
\sqrt{(1-c_2^2-c_4^2)}&\ketbra{0_{a2}0_{a1}}{0_{a2}1_{a1}}\Bigg\}
\end{split}
\end{equation}
\begin{equation}
\begin{split}
\mathcal{A}_{12}^{NB}\;\sim\; \Bigg\{\;
\cos{(\frac{\pi}{2M})}^M&\ketbra{1_{a2} 0_{a1}}{1_{a2} 0_{a1}},\\
\sqrt{(1-\cos{(\frac{\pi}{2M})}^{2M})}&\ketbra{0_{a2}0_{a1}}{1_{a2}0_{a1}},\\
\ketbra{0_{a2}0_{a1}}{0_{a2}1_{a1}}, &\ketbra{0_{a2}0_{a1}}{0_{a2}0_{a1}}\;\Bigg\}
\end{split}
\end{equation}
where coefficients $c_1$,$c_2$, $c_3$ and $c_4$ are functions of $M$ and $N$, with $c_2$ and $c_3$ going to 1, and $c_1$ and $c_4$ going to zero, as $N$ and $M$ go to infinity. This means one run (of $M$ outer cycles of $N$ inner cycles each) acts as a perfect optical switch in this limit, turning $H$ to $V$ (and vice-versa) if Bob blocks, and implementing identity if he doesn't.
\end{document}